\documentclass[referee,a4paper,12pt,traditabstract]{jswsc} 

\usepackage{graphicx}
\usepackage{txfonts}
\usepackage{epstopdf}
\usepackage[authoryear,round]{natbib}
\usepackage[backref]{hyperref}
\usepackage{url}

\hypersetup{colorlinks=true,citecolor=cyan,urlcolor=cyan,linkcolor=blue}

\begin{document}

   \title{Prediction of even and odd sunspot cycles}

   \titlerunning{Prediction of sunspots}
   \authorrunning{Asikainen and Mantere}

   \author{Timo Asikainen\inst{1} \and Jani Mantere\inst{1}}

\institute{Space physics and astronomy research unit, University of Oulu, PO Box 3000, 90014 Oulu, Finland\\
\email{\href{mailto:timo.asikainen@oulu.fi}{timo.asikainen@oulu.fi}}}

%%   \date{Received September 15, 1996; accepted March 16, 1997}

  % \abstract{}{}{}{}{}        %% uncomment if structured abstract is desired
 %% 5 {} token are mandatory
 
\abstract{Here we study the prediction of even and odd numbered sunspot cycles separately, thereby taking into account the Hale cyclicity of solar magnetism. We first show that the temporal evolution and shape of all sunspot cycles are extremely well described by a simple parameterized mathematical expression. We find that the parameters describing even sunspot cycles can be predicted quite accurately using the sunspot number 41 months prior to sunspot minimum as a precursor. We find that the parameters of the odd cycles can be best predicted with maximum geomagnetic $aa$ index close to fall equinox within a 3-year window preceding the sunspot minimum. We use the found precursors to predict all previous sunspot cycles and evaluate the performance with a cross-validation methodology, which indicates that each past cycle is very accurately predicted. For the coming sunspot cycle 25 we predict an amplitude of 171 $\pm$ 23 and the end of the cycle in September 2029 $\pm$ 1.9 years. We are also able to make a rough prediction for cycle 26 based on the predicted cycle 25. While the uncertainty for the cycle amplitude is large we estimate that the cycle 26 will most likely be stronger than cycle 25. These results suggest an increasing trend in solar activity for the next decades.}

\maketitle

%\keywords{Sunspots, Statistics; Solar Cycle, Observations; Solar Cycle, Models; Magnetosphere, %Geomagnetic Disturbances}
%\maketitle

%-------------------------------------------------
\section{Introduction}

Prediction of the sunspot number has been an everlasting interest in the space science community since the discovery of the sunspot cycle by \cite{Schwabe1844}. Sunspot number is an indirect indicator of many different solar phenomena, e.g., total and spectral solar radiation \citep[e.g.][]{Krivova2011,Frohlich2012}, coronal mass ejections \citep[e.g.][]{Richardson2012}, solar flares and magnetic active regions \citep[e.g.][]{Toriumi2017}. Its cyclic variation can even be used as a pacemaker to time different aspects of solar activity, solar wind and resulting geomagnetic variations \citep{Chapman2021,Leamon2022}. Therefore, there is considerable practical interest in predicting the evolution of future sunspot cycle(s). This is especially true in today’s technological society where space hazards pose a significant threat, e.g., to satellites, communications and electric grids on ground \citep[e.g.][]{Lanzerotti2001}. Another interest for predicting sunspots arises from the relatively recently recognized influences of variable solar radiation and solar wind activity on Earth’s climate system \citep{Gray2010,Ward2021,Salminen2020}.

Over the period of about last 100 years a vast array of different methods ranging from statistical methods to intensive physical simulations have been developed for predicting sunspots. As an unbiased introduction to all of them would be a futile effort here, the interested reader is referred to several excellent reviews on the subject by \cite{Hathaway2009}, \cite{Pesnell2012} and \cite{Petrovay2020}. According to the classic solar 
dynamo theory poloidal solar magnetic field in the solar minimum gets stretched to the toroidal magnetic field of the 
next cycle, which then produces sunspots and magnetic active regions that ultimately make up the poloidal field in the next solar minimum \citep{Parker1955,Babcock1961,Leighton1969,Charbonneau2020}. Physically motivated solar cycle predictions
are based on numerical modeling of the solar dynamo process and the transport of magnetic flux on the solar surface \citep{Charbonneau2020,Nandy2021,Karak2023}. However, some of the most successful, yet much simpler prediction methods are based on precursors that serve as indicators for the strength of the coming solar cycle. 
The so called polar field precursor methods have found a good correlation between the magnetic field observed at the solar polar region up to a few years before the sunspot minimum and the amplitude of the next sunspot cycle \citep{Schatten1978,Petrovay2020,Kumar2021,Kumar2022}. 
As the polar field reflects the strength of the poloidal phase of the solar magnetic field the precursor methods are deeply rooted in the core idea of the dynamo theory. 

Because the polar field has been systematically measured only since 1970s some longer running proxy measures for the polar field have also been used. 
The most successful polar field proxies are based on geomagnetic activity measures close to the solar minimum \citep[e.g.][]{Ohl1966,Du2009}.  
It has been shown that certain geomagnetic activity indices, e.g., the $aa$ index correlate quite well with the 
open solar magnetic flux carried by the solar wind \citep[e.g.][]{Lockwood1999,Lockwood2014}.
Close to the solar minimum the geomagnetic activity is dominantly driven by
high speed solar wind streams \citep[e.g.][]{Richardson2012}, which emanate from coronal holes that eventually form the solar polar field 
in the declining phase of the cycle \citep{Krieger1973,Bame1976}.
During these times the geomagnetic activity is most directly connected to the polar field
and, thereby acts as a good proxy for the amplitude of the next sunspot cycle.

In accordance with the solar dynamo theory the sunspot cycle prediction methods often consider only the predictability of the cycle based on the previous cycle.
Most of these methods do not typically take into account the 22-year Hale cycle of solar magnetism, 
which is well known in the solar cycle \citep[e.g.][]{Gnevyshev1948,Takalo2018,Leamon2022} and geomagnetic phenomena \citep[e.g.][]{Chernosky1966,Takalo2021,Chapman2021}. 
However, some recent studies have considered the even/odd cycle parity in the context of sunspot cycle prediction \citep[e.g.][]{Du2020b, Kakad2021, Penza2021,Du2022b,Nagovitsyn2023}.
Here we study the prediction of sunspot cycles accounting for the 22-year Hale cycle by considering
the differences in the even and odd numbered sunspot cycles. In Section 2 we first present our data and then in Section 3 proceed to 
show that the time evolution and shape of all sunspot cycles are extremely well described by a simple parameterized mathematical expression. 
In Section 4 we discuss the mutual dependencies of the parameters describing the sunspot cycles and in Section 5 we show that the parameters can be predicted using precursors found partly from past sunspot number values and partly from geomagnetic activity, which is often used as a precursor proxy for solar polar magnetic field in the sunspot minimum. 
Most importantly, though, we find that even and odd sunspot cycles obey different statistics therefore implying Hale cyclicity 
in their predictability. Separately these statistical relations are stronger and more significant than those based on combining even and odd cycles together. 
Using these statistics we construct in Section 6 a new method to separately predict even and odd sunspot cycles and apply it to the coming solar cycle 25. 
We also find an interesting connection between consecutive odd-even
sunspot cycle pairs, which allows us to make rough early predictions for cycle 26 as well. In Section 7 we discuss the results and give our conclusions.

\section{Data\label{sec:data}}
In this work we use monthly values of the version 2 Sunspot Number (SSN) obtained from the \cite{sidc}.
At the time of writing this paper the SSN v2 series covers time from 1749 to October 2022 therefore covering full sunspot cycles 1-24 and
about two years from the start of sunspot cycle 25. The monthly values of SSN were first smoothed with a 13-month window, 
where the first and last month are given a weight of 0.5 and all other months a weight of 1. 
Using this smoothed SSN curve we identified the times of sunspot minima and maxima.

In addition to the sunspot data we use in this work the geomagnetic $aa$ index. Recently \cite{Lockwood2018a,Lockwood2018b} presented a homogenized 
version of the $aa$ index, which inter-calibrates the observations of the different stations used to compose the $aa$ index. They also 
correct the index for the secular drift in the location of the auroral oval (due to secular changes in Earth's magnetic field) in relation 
to the observing stations. The homogenized $aa$ index was obtained from the supplementary data of \cite{Lockwood2018b}, which offers the 3-hourly values for years 1868 to 2017.

We first extended this series forward in time from Jan 2018 to Dec 2021 by calculating the monthly means of the homogenized $aa$ index and by calibrating 
the raw $aa$ index (obtained from ISGI: http://isgi.unistra.fr/) against the corresponding homogenized values.
The calibration was found by fitting a regression line to the logarithmic monthly averaged homogenized $aa$ ($aa_H$) and raw ($aa$) values
using the data between 1980 and 2017 when the $aa$ index is based on the latest pair of observatories (Hartland in England and Canberra in Australia).
The best fit line was 

\begin{equation}
\label{eq_aa_cal1}
\log(aa_H) = 1.049(\pm0.004) \times \log(aa) - 0.257(\pm0.011).
\end{equation}

The uncertainties of the fit parameters have been indicated in parentheses.
Note that this scaling uses logarithmic values, since in logarithmic scale the residuals of the fit are closely homoscedastic (constant variance) while in linear scale they display large heteroscedasticity thereby compromising the basic assumptions of the least-squares fit. 
The 3-hourly raw $aa$ values since Jan 2018 were then scaled with Eq. \ref{eq_aa_cal1} and the 
resulting data was appended to the homogenized $aa$ index time series. 

We also extended the $aa$ index backward in time using the daily magnetic declination based $Ak(D)$ indices recorded at the Helsinki geomagnetic observatory since 1844 \citep{Nevanlinna2004} and obtained from https://space.fmi.fi/MAGN/K-index/. These values have previously been used successfully to extend the $aa$ index time series backward from 1868 to 1844 \citep[e.g.][]{Lockwood2014}.

Unlike \cite{Lockwood2014}, who used annual averages, we calibrated here the daily $Ak(D)$ values against the simultaneous homogenized $aa$ index values by \cite{Lockwood2018b}
for the overlapping time period from 1.1.1868 to 31.12.1879, where the $Ak(D)$ data is rather continuous. Starting from 1880 the $Ak(D)$ data series has large data gaps.
We found that the daily $Ak(D)$ values can be scaled to the corresponding daily homogenized $aa$ index values by the following equation
\begin{equation}
    \log(aa_H) = 1.199(\pm0.012) \times \log(Ak(D)+5~\textnormal{nT}) - 0.97(\pm0.04).
\end{equation}
Also here the logarithmic scale ensures homoscedasticity of the fit residuals.
The correlation between the scaled $Ak(D)$ values and the homogenized $aa$ index values is 0.84 indicating a rather reliable scaling. 

\begin{figure*}[htbp]
\centering
\includegraphics[width=\textwidth]{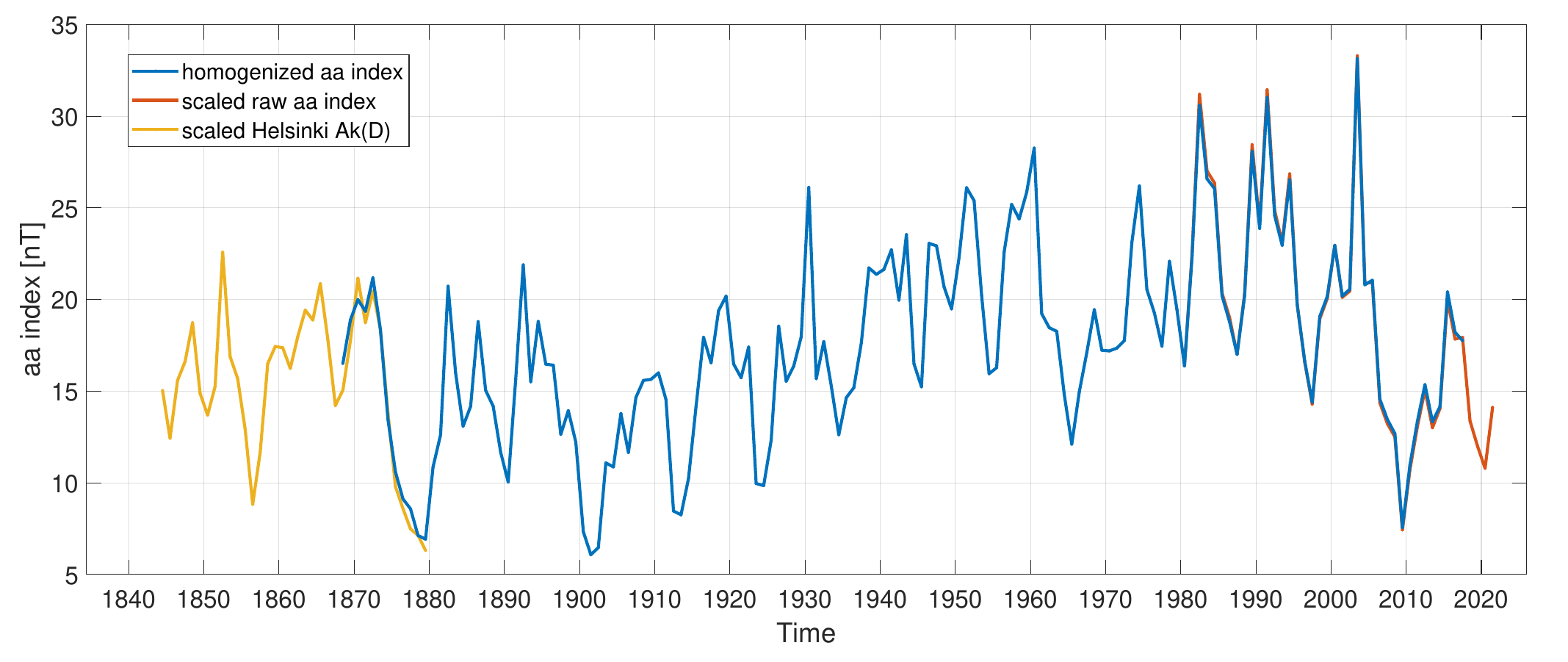}
\caption{Annual averages of homogenized $aa$ index (blue), scaled raw $aa$ index (red) (shown since year 1980) and scaled Helsinki $Ak(D)$ values (shown for years 1844-1879).\label{fig_aa1}}
\end{figure*}
Figure \ref{fig_aa1} shows the annual averages of homogenized $aa$ index (blue), scaled raw $aa$ index (red) (shown since year 1980) and scaled Helsinki $Ak(D)$ values (shown for years 1844-1879). The times when the different curves overlap indicate a very good correspondence between the different datasets. The extended $aa$ index time series is formed 
by the scaled $Ak(D)$ data from 1844-1867, by the homogenized $aa$ index from 1868-2017 and by the scaled raw $aa$ index since 2018. For the purposes of this study we will use
monthly averages of the extended $aa$ index composite.

\section{Parameterization of the sunspot cycle}

Sunspot cycles are often fitted with a parameterized curve \citep[e.g.][]{Stewart1938,Hathaway1994,Volobuev2009}. 
\cite{Stewart1938} showed that the sunspot cycles could be described roughly by curves of the form $c(t-t_0)^a e^{-b(t-t_0)}$, where $a,b,c$ and $t_0$ are free parameters. \cite{Hathaway1994} used a slightly modified version
\begin{equation}
f(t) = \frac{a(t-t_0)^3}{\exp((t-t_0)^2/b^2)-c}
\label{eq_hathaway}
\end{equation}
and fitted this model curve to sunspot cycles 1-21. They also showed that many of the parameters correlated rather well with each other thereby offering a possibility to reduce the effective number of parameters in the curve down to one free parameter. Many studies have since used similar parameterizations.
However, to be useful in predicting the sunspot cycles the parameters of the future cycle should be predicted 
by some means. Several studies have found relatively good correlations between different precursors and the amplitude 
of sunspot cycle or the maximum SSN during the cycle. The parameterizations described above (e.g., Eq. \ref{eq_hathaway}) may not be optimal in light
of these correlations, because for those the amplitude (maximum) of the sunspot cycle depends on a combination of several parameters.

Therefore, we formulated a new parameterization for the sunspot curve, where the parameters are perhaps better interpreted
in terms of well known solar cycle properties (amplitude, rise time, asymmetry etc.). We use here an asymmetric Gaussian curve of form
\begin{equation}
    f(t) = A\exp\left(-\frac{(t-B)^2}{(C\times g(t,B))^2}\right),
    \label{eq_model}
\end{equation}
where $A$ is the sunspot cycle maximum, $B$ is the time of the sunspot maximum measured from the sunspot minimum beginning
the cycle (i.e., $B$ is the cycle rise time), $C$ is the time scale of the rising phase (cf. standard deviation of a Gaussian) and function $g(t,B)$ is defined
as
\begin{equation}
\label{eq_model2}
g(t,B) =\left\{\begin{array}{cc}
             1, & \textnormal{if } t\leq B,\\
             D, & \textnormal{if } t> B.\end{array}
             \right.
\end{equation}
Therefore the parameter $D$ appearing in $g(t,B)$ describes the asymmetry of the time scales in the declining and rising phases.
The more positive the $D$ is the longer the declining phase is compared to the rising phase.

We then fitted the Eq. \ref{eq_model} to the 13-month smoothed SSN of each sunspot cycle separately. Each cycle was defined 
from the time of the sunspot minimum to the next minimum.
The 4-parameter fit was done with the non-linear Levenberg-Marquardt
optimization implemented in Matlab software. Although the consecutive sunspot cycles are known to be overlapping (so that the cycle already starts before the minimum SSN time) the fitting of our parametric model is largely dependent on the whole cycle and results are not strongly sensitive to the data around sunspot minima. This was tested either by leaving out up to 1 year of data from the beginning of the cycles or by extending the previous fitted cycle and subtracting it from the next cycle. In either case the fitted parameter values remained practically the same.

\begin{figure*}[htbp]
\centering
\includegraphics[width=\textwidth]{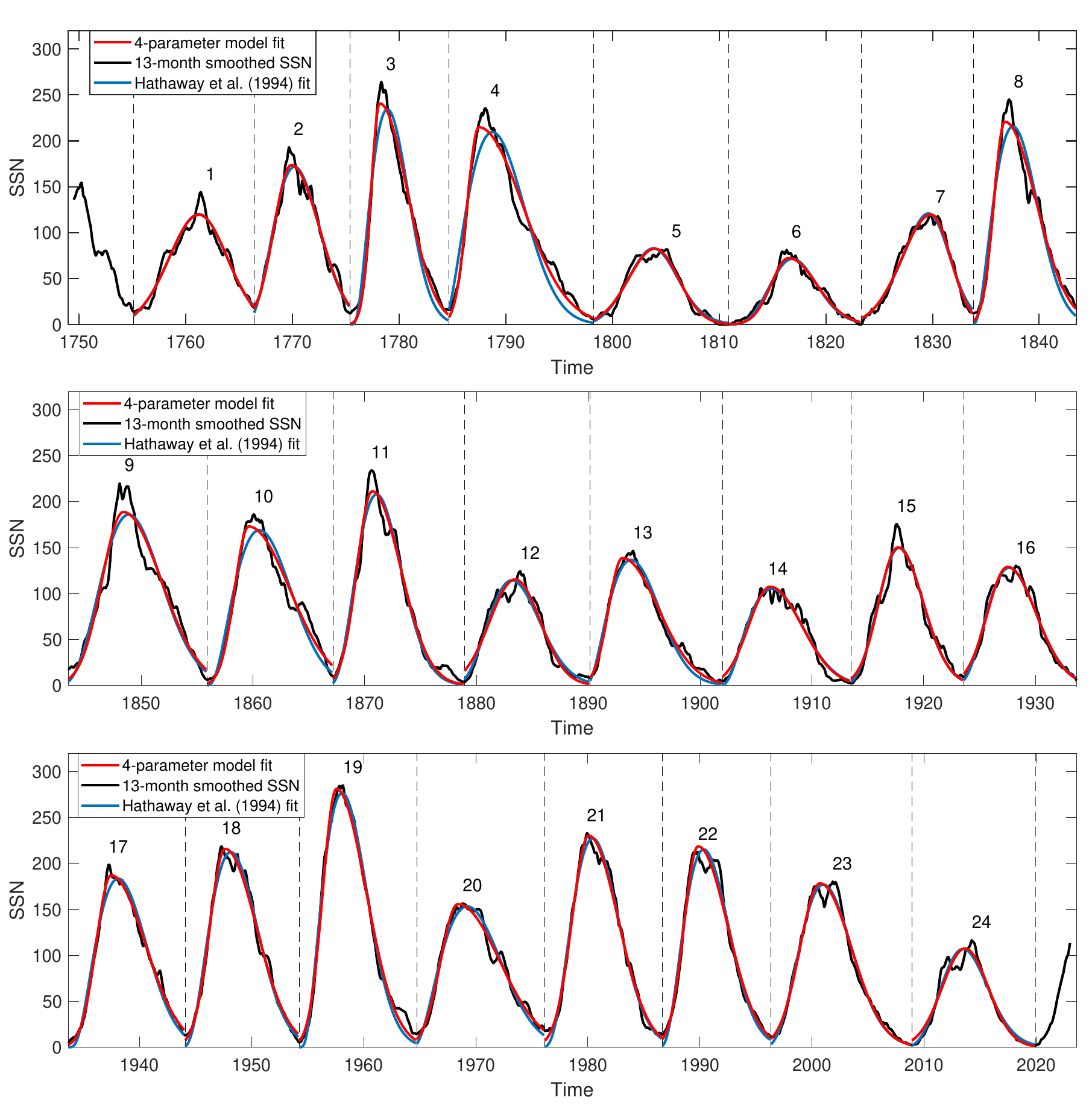}
\caption{Time series of 13-month smoothed sunspot number (black), parameterized fits for each sunspot cycle (red curves) and the fit by \cite{Hathaway1994} for comparison (blue curves). 
The sunspot cycle numbers are indicated on the plot as well. The top panel shows cycles 1-8, middle panel cycles 9-16 and bottom panel cycles 17-24. }
\label{fig1}
\end{figure*}

Figure \ref{fig1} shows the time series of the 13-month smoothed SSN in black, the 4-parameter
model fits for cycles 1-24 in red and the \cite{Hathaway1994} fit of Eq. \ref{eq_hathaway} in blue. One can see that the fitted curves describe all the individual cycles reasonably well,
although some of the detailed structure in the SSN cycles cannot be described by a smooth asymmetric Gaussian. Such structures are for example the very sharp-peaked cycles like cycles 1-4 and 8-11 and the often seen double peaks (see, e.g., \cite{Karak2018}), which are quite prominent, e.g., in cycles 22-24.
However, the rising and declining phases of each cycle are well captured by the fit and the cycle amplitudes are quite close to the 
real cycle amplitudes. The average $R^2$-value of the 4-parameter fit over all cycles is 0.973, indicating that over 97\% of the variability 
in the SSN cycles is captured by the model curves. The average root-mean-squared error for all the cycles is 8.7 and there is no 
statistically significant difference in this between even and odd numbered sunspot cycles. We also note that for most cycles 
the asymmetric Gaussian function used here is very close to the function (Eq. \ref{eq_hathaway}) used by \cite{Hathaway1994}. However, in some cycles (3, 4, 8, 10) there is a clear difference with the asymmetric Gaussian providing a somewhat better fit.

\section{Relationships between cycle parameters}

\begin{figure*}[htbp]
\centering
\includegraphics[width=\textwidth]{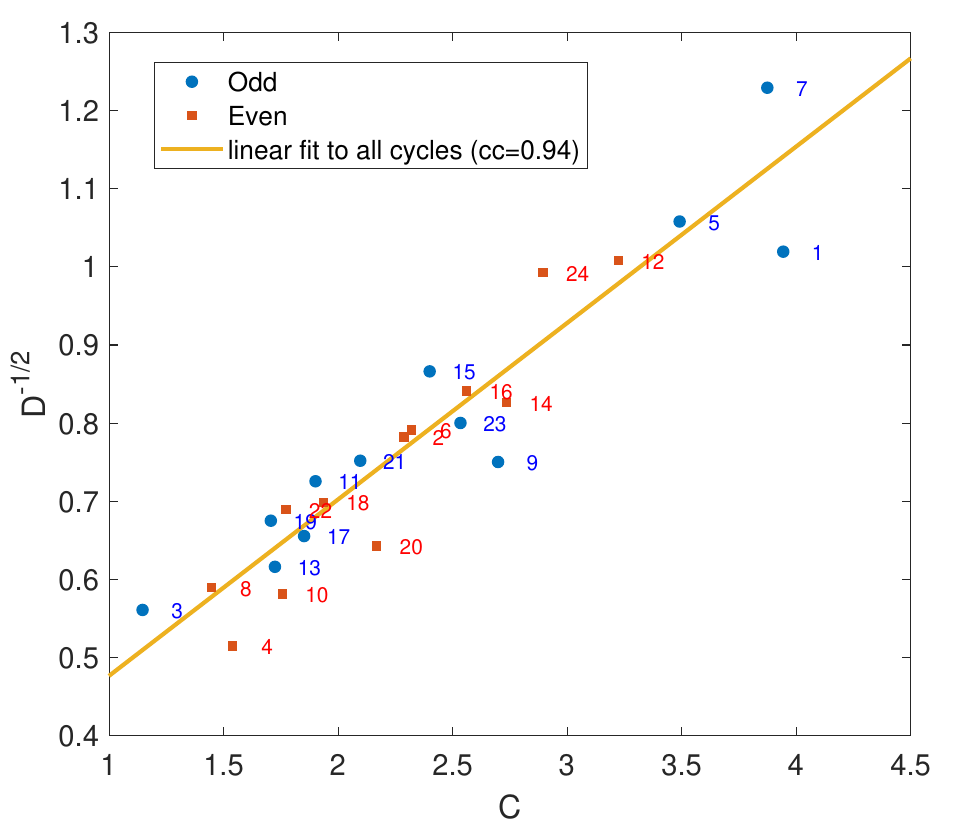}
\caption{Relationship between the $C$ and $D^{-1/2}$ parameters of the 4-parameter fits. Blue dots (red squares) indicate even (odd) numbered cycles.
The cycle numbers have been further indicated beside all points. The yellow line depicts the linear fit to all the cycles having a correlation 
coefficient of 0.94 (p-value is $10^{-11}$)}
\label{fig2}
\end{figure*}

Let us next consider the relationships between the fitted values of the four parameters over all cycles. Figure \ref{fig2} shows as scatter plot 
the relationship between parameters $C$ and $D^{-1/2}$. In the figure the odd numbered cycles have been indicated with blue dots and even 
numbered cycles with red squares. One can see that all the cycles depict quite a strong correlation between $C$ and $D^{-1/2}$ and that there
is no significant difference between the odd and even cycles. The correlation coefficient between $C$ and $D^{-1/2}$ over all the cycles is 0.94
and it is highly significant (p-value is $10^{-11}$). This relationship indicates that cycles with a steep rising phase (i.e., small $C$) 
have a relatively more gradual (i.e., large $D$ and small $D^{-1/2}$) declining phase and vice versa. The correlation in Figure \ref{fig2}
is therefore a manifestation of the well known property of sunspot cycles that small amplitude cycles tend to be more symmetric about 
the sunspot maximum time than large amplitude cycles \citep{Waldmeier1968}. 

For our purposes the tight relationship between $C$ and $D$ allows us to eliminate the parameter $D$ from the sunspot model curve and 
thereby reduce the number of free parameters to 3. The parameter $D$ is therefore replaced in Eqs. \ref{eq_model} and \ref{eq_model2} by 
expression
\begin{equation}
    D = \left(0.25(\pm0.05) + 0.226(\pm0.02)\times C\right)^{-2},
    \label{eq_D}
\end{equation}
which corresponds to the linear fit (yellow line) in Figure \ref{fig2}.

After replacing $D$ with Eq. \ref{eq_D} we repeated the fitting procedure for each sunspot cycle, but now using the remaining 3-parameter model.
The correlation between $C$ and $D^{-1/2}$ is so high that the 3-parameter model is roughly equally good as the 4-parameter model in describing each sunspot cycle. 
The average $R^2$-value for the 3-parameter model is 0.959 and therefore not much smaller than 0.973 for the 4-parameter model.
Note also that the elimination of $D$ from the model does not significantly alter the values of the remaining parameters $A$, $B$ and $C$. 
The correlations of the corresponding parameters of the 4-parameter and 3-parameter fits exceed 0.97.

\begin{figure*}[htbp]
\includegraphics[width=\textwidth]{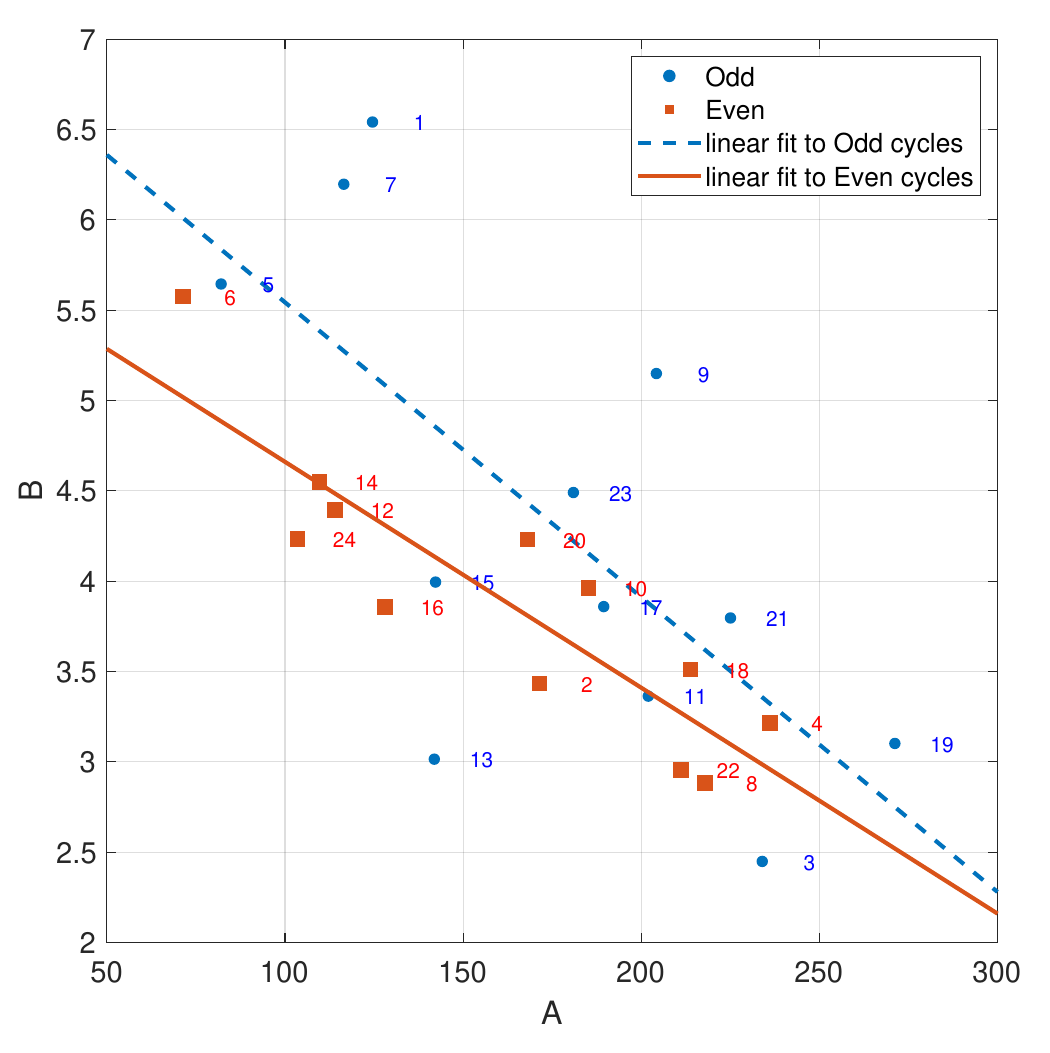}
\caption{Relationship between the $A$ and $B$ parameters of the 3-parameter fits. Blue dots (red squares) indicate even (odd) numbered cycles.
The cycle numbers have been further indicated beside all points. The blue and red lines depict the linear fits to odd and even cycles respectively.}
\label{fig3}
\end{figure*}

Figure \ref{fig3} shows the relationship between the parameter $B$ (time of SSN maximum counted from SSN minimum in years) and cycle amplitude $A$. 
One can see that there is a general anti-correlation between $A$ and $B$. This anti-correlation between the cycle rise time and amplitude has been 
long known as the Waldmeier effect \citep{Waldmeier1935}. 
Here one can see that this anti-correlation is somewhat stronger (cc = -0.88, p-value = $10^{-4}$) in even cycles than in odd cycles (cc = -0.69, p-value = 0.014). The correlation between $A$ and $B$ found here is
clearly higher than, e.g., the correlation between SSN maximum and cycle rise time found by \cite{Karak2011} (cc=-0.5). It therefore appears that 
the parameters of the fitted SSN model curve indicate the Waldmeier effect more robustly than the exact SSN data. This is likely because
of the sensitivity of the Waldmeier effect to the of timing and height of the sunspot maximum, which can be difficult to determine if the cycle has multiple peaks \cite{Karak2011}.

The linear fits to even and odd cycles depicted by the red and blue lines in Figure \ref{fig3} are given by equations
\begin{eqnarray}
\label{eq_BA_even} B_{\textnormal{even}} &=& 5.9(\pm0.4) - 0.013(\pm0.003)\times A_{\textnormal{even}} \\
\label{eq_BA_odd} B_{\textnormal{odd}} &=& 7.2(\pm1.0) - 0.016(\pm0.006)\times A_{\textnormal{odd}}.
\end{eqnarray}
The correlation coefficients and the slopes/intercepts of the best fit regression lines 
are different in the even and odd cycles only with a weak statistical significance (the p-values for the differences exceed 0.12). 
However, the mean squared errors of the linear fits to the even and odd cycles are highly significantly different 
(0.14 for even cycles and 1.01 for odd cycles and the p-value for the difference is 0.003).
This result indicates that the Waldmeier effect is more strongly valid for even cycles than for odd cycles. 

\begin{figure*}[htbp]
\centering
\includegraphics[width=\textwidth]{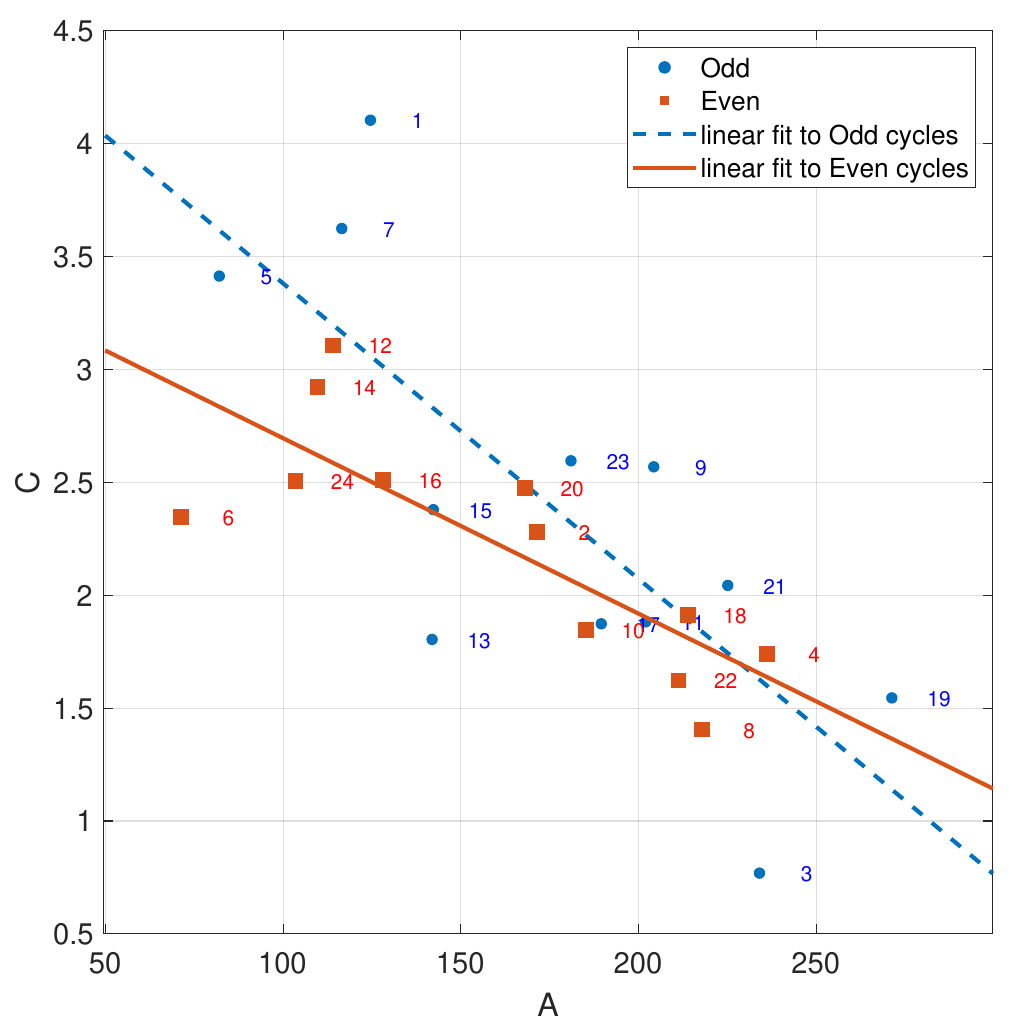}
\caption{Relationship between the $A$ and $C$ parameters of the 3-parameter fits. Blue dots (red squares) indicate even (odd) numbered cycles.
The cycle numbers have been further indicated beside all points.  The blue and red lines depict the linear fits to odd and even cycles respectively.}
\label{fig4}
\end{figure*}

Figure \ref{fig4} shows the relationship between the final parameter $C$ (time scale of rising phase) and cycle amplitude $A$.
Here one can also see a general anti-correlation, which is yet another manifestation of the Waldmeier effect. 
Actually, there is also a quite good correlation (cc = 0.91, p-value = $8\times10^{-10}$) between the $B$ and $C$ parameters, which explains the two manifestations of the Waldmeier effect.
The linear fits to even and odd cycles depicted by the red and blue lines in Figure \ref{fig3} are given by equations
\begin{eqnarray}
\label{eq_CA_even} C_{\textnormal{even}} &=& 3.5(\pm0.4) - 8(\pm2)\times 10^{-3}\times A_{\textnormal{even}} \\
\label{eq_CA_odd}  C_{\textnormal{odd}} &=& 4.7(\pm0.7) - 1.3(\pm0.4)\times 10^{-2} \times A_{\textnormal{odd}}.
\end{eqnarray}

The conclusions about the 
differences between even and odd cycles are the same as for $B$ vs. $A$ relation in Figure \ref{fig3}. I.e., the linear relationships 
or the correlations are only weakly statistically significantly different, but the difference in the mean squared errors is highly significant (p-value is 0.038).

Overall the results in Figures \ref{fig3}-\ref{fig4} indicate that the cycle amplitude $A$ could further be used to reduce the number of parameters in the sunspot cycle fit. Moreover, there are indications that the accuracies of these fits are significantly different in even and odd cycles.

\section{Precursors for cycle parameters}

\begin{figure*}[htbp]
\includegraphics[width=\textwidth]{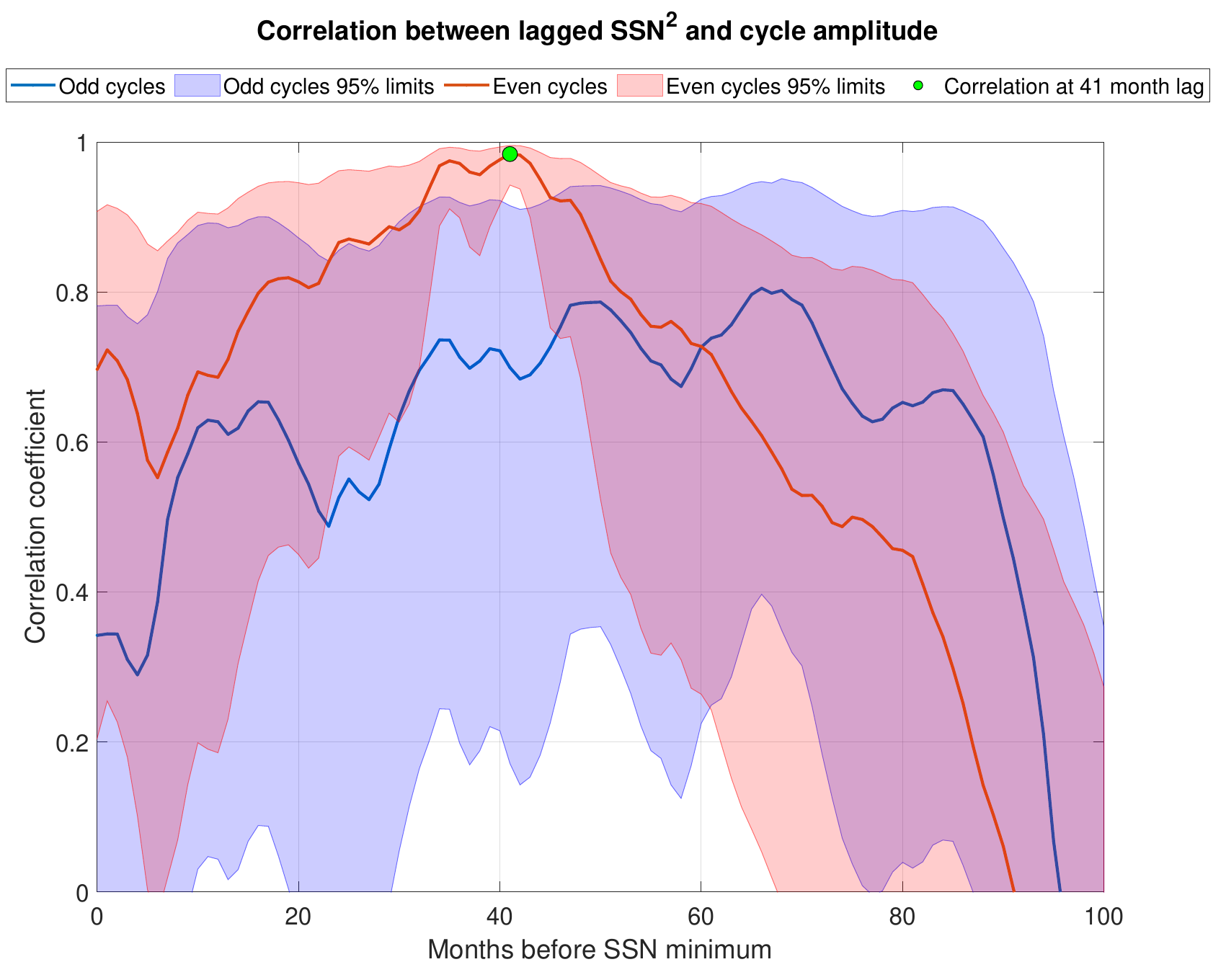}
\caption{Correlation coefficient between cycle amplitude $A$ and lagged 13-month smoothed sunspot number (SSN$^2$) as a function of the time lag in months before the SSN minimum.
Blue (red) curve indicates the odd (even) cycles and the correspondingly colored regions indicate the 95\% confidence limits for the correlation coefficients.
The green dot indicates the maximum correlation at 41 months before the SSN minimum for the even cycles.}
\label{fig5}
\end{figure*}

Based on the above relations we could further reduce the SSN model curve
to a one parameter model. However, each simplification of the model makes
it less flexible and decreases the model accuracy. Therefore it is useful to 
first consider whether the cycle parameters $A$, $B$ and $C$ can be directly predicted by some suitable precursor. \cite{Cameron2007} showed that the solar activity level three years before
the sunspot minimum starting the cycle is a relatively good predictor for the amplitude of the cycle. This result for sunspot numbers was shown by \cite{Petrovay2020} (his Figure 6),
who found a correlation of 0.8 between maximum sunspot number of the cycle and the sunspot number taken 3 years before the sunspot minimum that starts the cycle.
A correlation of 0.8 implies that only about 62\% variability in cycle amplitudes could be explained by the past sunspot number, thereby offering a rather mediocre 
accuracy in predicting the coming cycle amplitudes as also \cite{Petrovay2020} mentions. However, still motivated by this result we calculated in Figure \ref{fig5} 
 the correlation coefficient between the cycle amplitude $A$ and the lagged sunspot number as a function of the time lag in months before the SSN minimum. Unlike in 
 past studies we did the calculation here separately for even and odd cycles and using the 2nd power of SSN (SSN$^2$), which we found to produce slightly larger correlations than SSN (the difference, however, is rather small and not statistically significant).
 
\begin{figure*}[htbp]
\includegraphics[width=\textwidth]{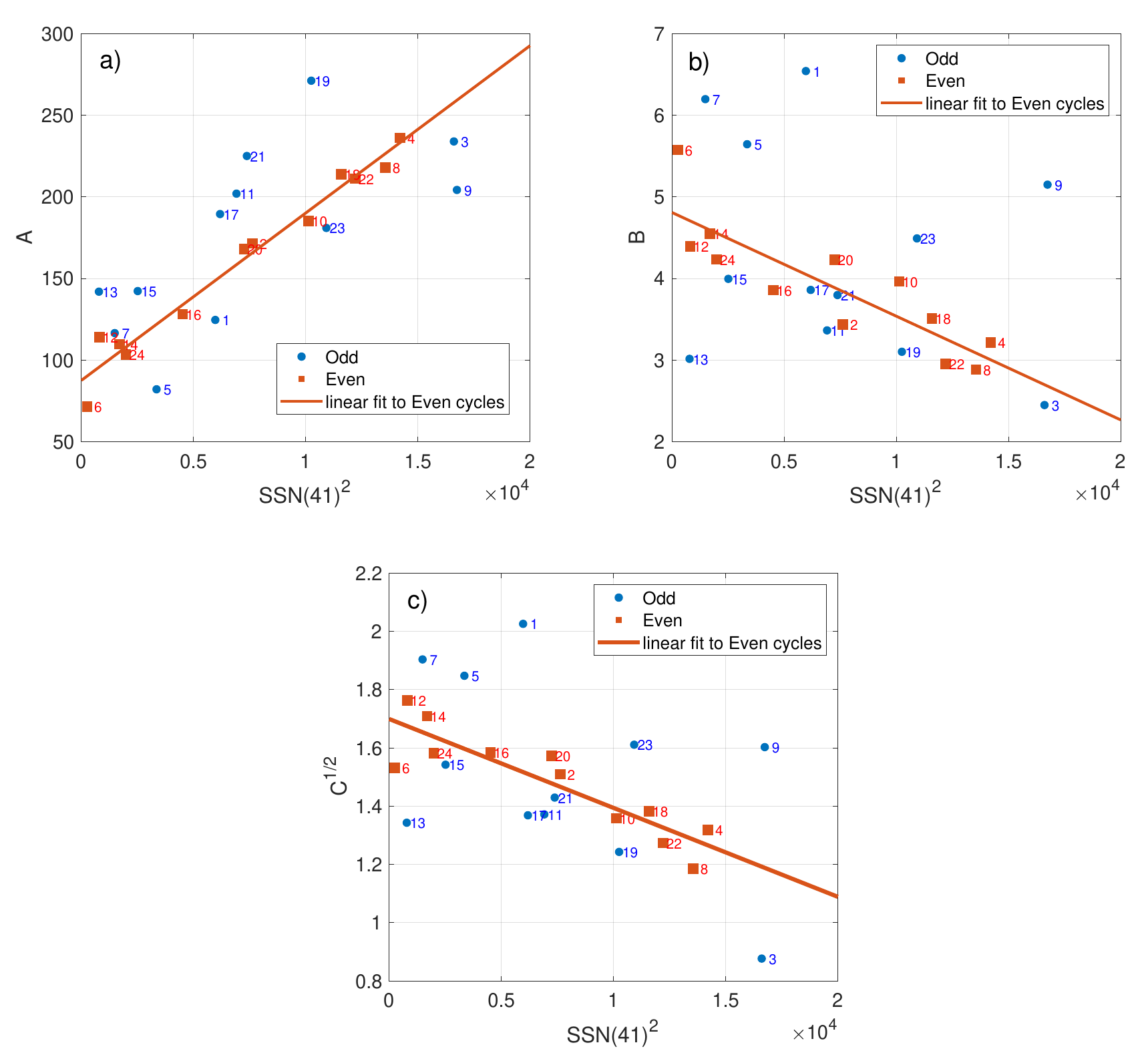}
\caption{Dependence of $A$, $B$ and $C$ parameters on 13-month smoothed sunspot number 41 months before the sunspot minimum (SSN(41)$^2$). 
Blue dots (red squares) indicate odd (even) numbered cycles.
The cycle numbers have been further indicated beside all points. The red lines depict the linear fits to even numbered cycles.}
\label{fig6}
\end{figure*}

One can see that for the even cycles the correlation is systematically higher than for the odd cycles. The difference between even and odd cycles is quite large, but because of the
rather low number of data points (only 12 even and odd cycles) the 95\% confidence limits remain rather large even when the correlation coefficient is high.
However, there is a location around the lag of 41 months (about 3.4 years) before sunspot minimum where the correlations do differ from each other statistically significantly. This lag of optimal correlation is quite close to the 3 years found by \cite{Cameron2007}. Note, however, that the good correlation is not specific to exactly the optimum of 41 months but is seen over a broad range of lags from 34 to 44 months.

We chose the lag of 41 months for a closer inspection in Figure \ref{fig6}, which displays all the three parameters $A$, $B$ and $C^{1/2}$ as a function of the sunspot number
taken 41 months before the sunspot minimum (SSN(41)). One can see that the SSN(41)$^2$ correlates quite well not only with cycle amplitude $A$, but with all the three parameters in the even cycles. 
In the odd cycles there is evident correlation too, but it is clearly much lower than for the even cycles due to the larger scatter.
In particular the SSN(41)$^2$ vs. cycle amplitude $A$ resembles the plot in Figure 6 of \cite{Petrovay2020}, but now reveals a large difference between even and odd cycles. Recently \cite{Du2020b} also found that the sunspot number 39 months before the sunspot minimum is a precursor for the maximum SSN of the next cycle and this relation is stronger for even cycles. A similar finding was done by \cite{Nagovitsyn2023}.

We shall later discuss the reasons for the better correlation in even cycles but for now we concentrate on the fact that SSN(41) can be used as a quite accurate predictor of 
the three cycle parameters for the even cycles. For cycle amplitude the correlation coefficient is 0.984 (p-value = $8\times 10^{-9}$) indicating that about 96.9\% of the variation in cycle amplitude may be predicted with SSN(41). For cycle rise time $B$ the correlation is slightly lower -0.86 (p-value = 0.0003) and for $C^{1/2}$ (time scale of rising phase) the correlation is -0.89 (p-value = $10^{-4}$).

The linear fits to the even cycles indicated by the yellow lines in Figure \ref{fig6} are given by equations
\begin{eqnarray}
\label{eq_even1} A_{\textnormal{even}} &=& 88(\pm 6) + 0.0103(\pm0.0006)\times \textnormal{SSN(41)}^2,\\
\label{eq_even2} B_{\textnormal{even}} &=& 4.8(\pm 0.3) - 1.3(\pm 0.3)\times 10^{-4} \times \textnormal{SSN(41)}^2,\\
\label{eq_even3} C_{\textnormal{even}} &=& \left(1.70(\pm 0.05) - 3.1(\pm 0.5)\times 10^{-5} \times \textnormal{SSN(41)}^2\right)^2.
\end{eqnarray}

\begin{figure*}[htbp]
\includegraphics[width=\textwidth]{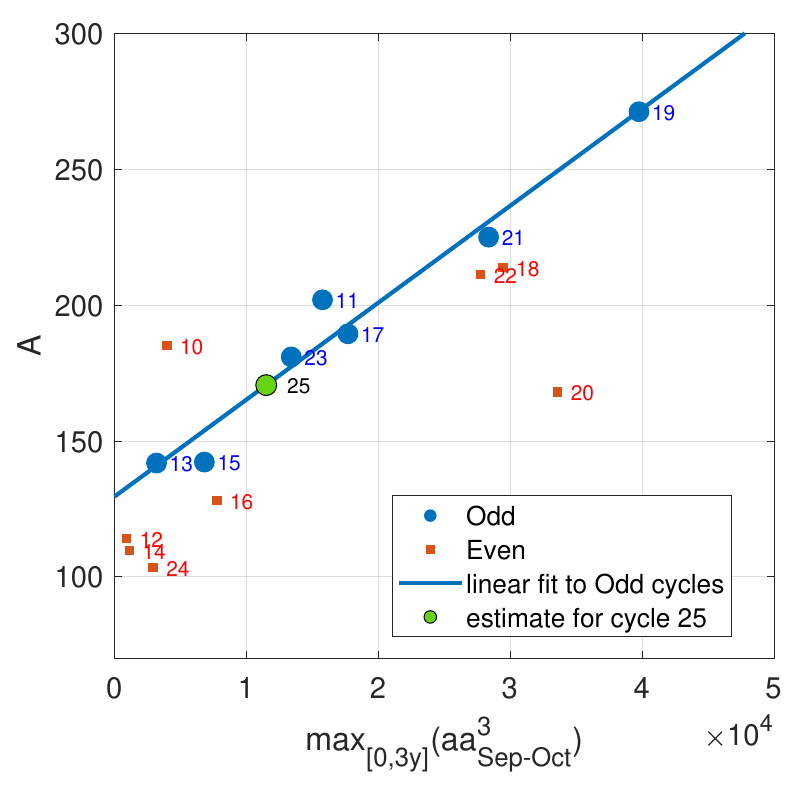}
\caption{Relationship between cycle amplitude $A$ and maximum September-October $aa^3$ from 
a 3-year window before SSN minimum. The blue points indicate odd cycles and red squares the even cycles. The blue line indicates the linear fit to the odd cycles. The green circle indicates the estimated amplitude for cycle 25.}
\label{fig7}
\end{figure*}

\begin{figure*}[htbp]
\includegraphics[width=\textwidth]{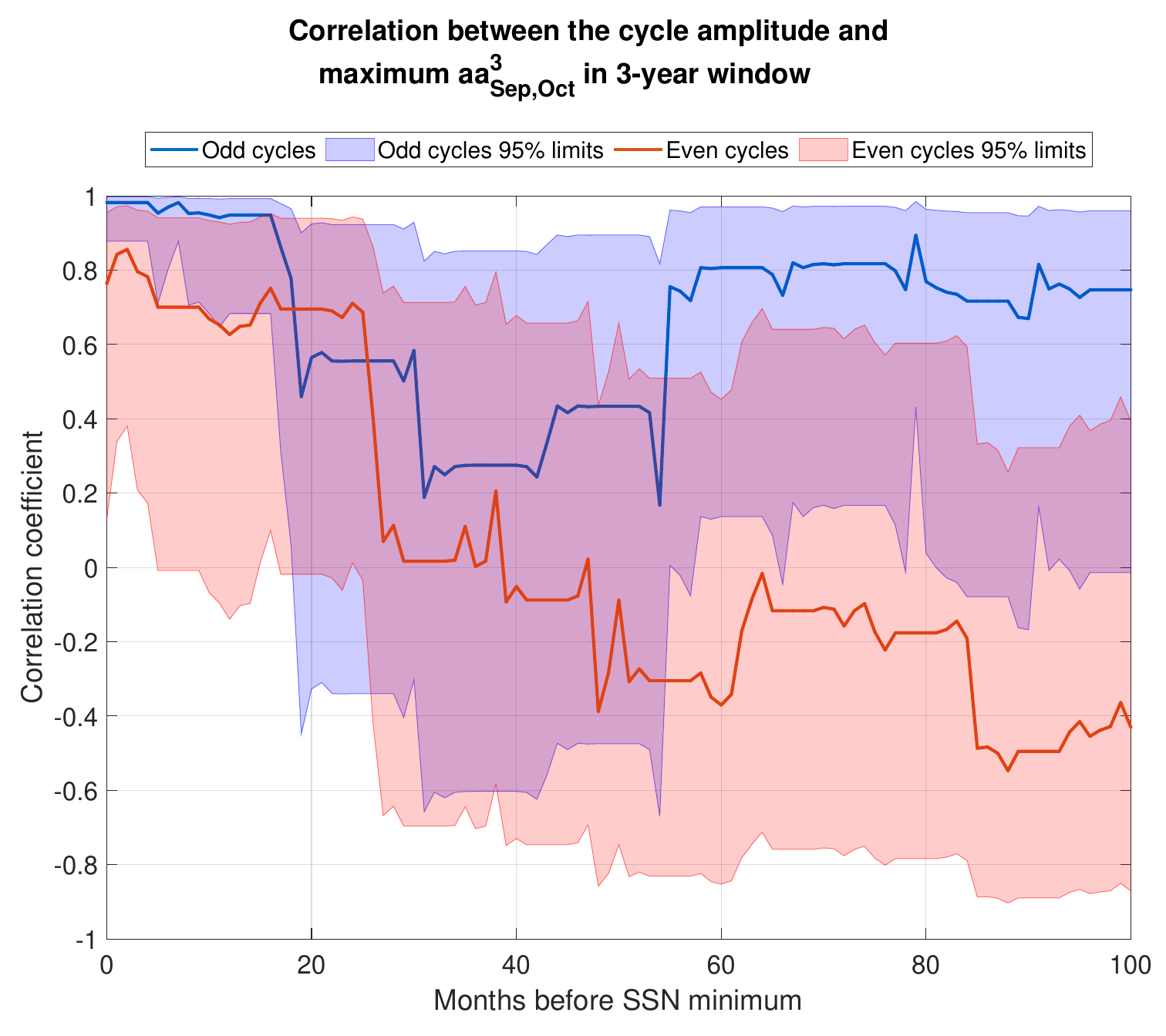}
\caption{Correlation coefficient between cycle amplitude $A$ and lagged maximum monthly September-October $aa^3$ within a 3 year window.
Blue (red) curve indicates the odd (even cycles) and the correspondingly colored regions indicate the 95\% confidence limits for the correlation coefficients.}
\label{fig8}
\end{figure*}

We also studied the correlation between cycle amplitude and geomagnetic activity, which is known to be a good 
precursor for the sunspot cycle amplitude. We tested various different quantities calculated from the extended homogenized geomagnetic $aa$ index.
We found that the best predictor for the cycle amplitude for odd sunspot cycles is given by the maximum average September-October $aa^3$ value
within the 3-year window extending backward from the sunspot minimum.
Figure \ref{fig7} shows the relationship between the cycle amplitude and this geomagnetic activity measure separately for the odd cycles (blue)
and even cycles (red). The yellow line indicates the best linear fit of 
\begin{equation}
 \label{eq_GMA_prec} A = 130(\pm7) + 0.0036(\pm 0.0004)\times \textnormal{max}_{[0,3\textnormal{y}]}\left(aa_{\textnormal{Sep-Oct}}^3\right).
\end{equation}
The linear correlation coefficient between the odd cycle amplitude and maximum $aa_{\textnormal{Sep-Oct}}^3$ is 0.981 (p $<9\times 10^{-5}$).
One can also see from Figure \ref{fig7} that the corresponding relation for the even cycles is much worse,
but still statistically significant (correlation 0.76, p = 0.027). It should be noted though, that trying different quantities calculated from 
the $aa$ index one can obtain higher correlations for the even cycles as well, but it seems that none of such correlations exceeds
the extremely good correlation between even cycle amplitude and SSN(41) found above. 

We also evaluated the correlation between the cycle amplitude and the geomagnetic precursor defined above as a function of 
time lag counted from the sunspot minimum preceding the cycle. I.e., instead of taking the maximum $aa_{\textnormal{Sep-Oct}}^3$ in a 3-year window ending at
the cycle minimum the window was shifted backward in time by varying lags. The resulting correlations are shown in Figure \ref{fig8} separately
for odd (blue) and even (red) cycles. One can see that the highest correlations for odd cycles are indeed found up to 16 months before the cycle minimum.
There is a drop in correlation between 16-55 months before minimum but the correlation then rises again and remains rather higher (about 0.8) 
between 55-100 months before the SSN minimum. Note however, that this does not imply a longer lead time for prediction, since the timing of the 
SSN minimum is still needed to determine the precursor. Figure \ref{fig8} also shows that the correlation for even cycles is clearly much lower than 
for the odd cycles at practically all lags. However, because of the somewhat limited number of data points the statistical 95\% confidence limits for
the correlations are rather large. Even though there are indications for systematically higher correlation for the odd cycles, 
there is no firm evidence that the correlations are statistically significantly different for the even and odd cycles.

The specific recipe of calculating the geomagnetic activity based precursor for odd cycles was found by trial and error, manually testing 
a few tens of different (rather random) possibilities. This raises the question of whether the found precursor is statistically significantly
better than some other, perhaps more common, choice like a mere the average $aa$ index in the year of sunspot minimum.
We tested the possibility of randomly obtaining a geomagnetic activity based precursor as good as the one found above.
This was done by generating $10^4$ random geomagnetic precursor time series by varying the calendar months from which $aa$ index is taken (2 random calendar months), the
length of the time window (randomly selected between 1 and 5 years) and the exponent assigned for the $aa$ index (randomly selected between 1 to 3). 
In addition we randomly varied whether we take the maximum of the yearly values within the time window (as we did for the chosen precursor) or the average. For each such randomly generated precursor
we calculated the correlation between the precursor and the following cycle amplitude as a function
of lag up to 11 years and found the maximum correlation. This procedure simulates the act of randomly
choosing a recipe for determining the precursor and finding the maximal correlation over all these lags. Finally, we randomly grouped the remaining correlations into groups of 10 values and determined the maximum correlation in each group. This simulated the act of testing 10 random precursors and choosing the 
one that gives the maximal correlation at some lag.
Now, comparing the correlation coefficient (0.981) for the precursor used above in Figures \ref{fig7} and \ref{fig8} to the maximal correlations of these randomly generated precursors in shows that there is only a probability of less than 4.5\% to randomly obtain a correlation higher than 0.981. 
This indicates that the recipe for the geomagnetic precursor we have chosen for the odd cycles 
is indeed statistically significant by 95\% significance level.
We also compared the geomagnetic precursor found above to a more commonly used geomagnetic precursor, which is the annual average of the $aa$ index at the solar minimum, for which the correlation with the following cycle amplitude is 0.834. The correlation for our precursor was 0.981 and the difference of it from 0.834 is statistically highly significant (p-value for the difference is about 0.0007).

\begin{figure*}[htbp]
\includegraphics[width=\textwidth]{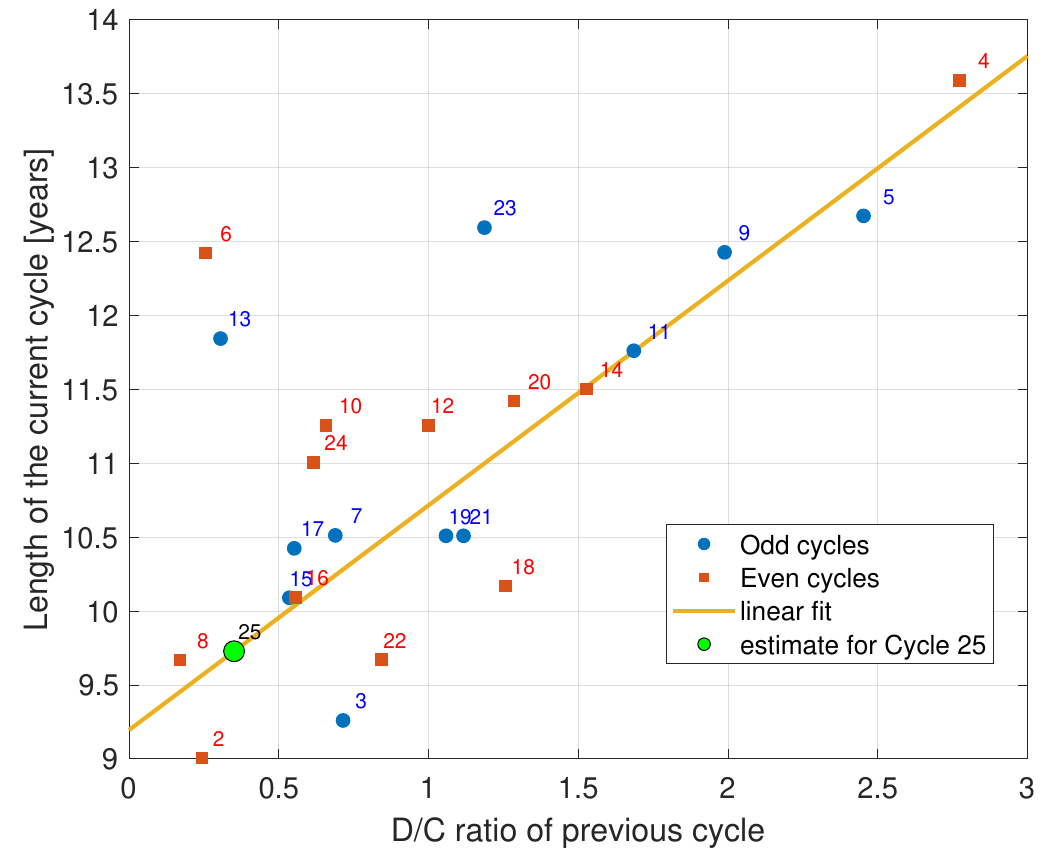}
\caption{Relationship between the cycle length and the $D/C$ ratio of the previous cycle. The numbers of each data point refer to the current cycle.
The blue points (red squares) correspond to odd (even) cycles and the green point to the estimated length of the cycle 25. The yellow line indicates the 
linear fit excluding the three cycles 6, 13 and 23.}
\label{fig8a}
\end{figure*}

The length of the sunspot cycle is also an interesting and important quantity. However, it is difficult to estimate from the predicted sunspot cycle curve
because typically the SSN does not drop to zero at the end of the cycle. We studied the association of the cycle length and the four cycle parameters,
but found no strong relationships, which would allow one to estimate the length of the cycle based on the parameters of the same cycle.
However, we found that the length of the cycle seems to significantly correlate with the ratio of $D$ and $C$ parameters (i.e., $D/C$) of the \textit{previous cycle}. 
This relationship is shown in Figure \ref{fig8a}. The correlation coefficient between $D/C$ of the previous cycle and length of the current cycle
is 0.66 (p $=5.7\times10^{-4}$) and is statistically very significant. There are three cycles (6, 13 and 23) which are clear outliers. Excluding these
cycles yields an even higher and more significant correlation of 0.88 (p $=2.6\times 10^{-7}$). We note that these correlations are both significantly higher than,
e.g., the correlation between the cycle length and the amplitude of the same cycle (about -0.5) \citep{Wolf1861,Petrovay2020}.

The yellow line in Figure \ref{fig8a} shows the linear fit excluding the cycles 6, 13 and 23. The equation for this fit is
\begin{equation}
    L_{i} = 9.2(\pm0.4) + 1.5(\pm0.3) \frac{D_{i-1}}{C_{i-1}},
    \label{eq_cycle_length}
\end{equation}
where $L_i$ is the length of cycle $i$ in years and $D_{i-1}$ and $C_{i-1}$ refer to the parameters of previous cycle $i-1$.

\section{Prediction of sunspot cycles}
\subsection{Cross-validated predictions for past cycles and prediction for cycle 25}
Based on the discussion above we can predict the sunspot cycle once the time of the sunspot minimum starting the new cycle is known.
We use the 3-parameter description ($A$, $B$ and $C$) for the sunspot cycle of Eqs. \ref{eq_model}-\ref{eq_model2}.
Parameter $D$ has been eliminated using Eq. \ref{eq_D}.
For the even sunspot cycles we directly estimate the three $A$, $B$ and $C$ parameters with SSN(41)$^2$ as shown in Figure \ref{fig6}.

For the odd cycles we first estimate the cycle amplitude $A$ from the geomagnetic precursor as shown in Figure \ref{fig7} 
and then estimate $B$ and $C$ using the linear relationships for the odd cycles given depicted in Figures \ref{fig3} and \ref{fig4} respectively.

Using this approach we predicted past even sunspot cycles starting from cycle 2 and each odd sunspot cycle starting from cycle 11.
We predicted each cycle separately with the so-called leave-one-out cross validation method. This means that when predicting the $i$:th cycle
all the fits between different parameters and precursor values discussed above were determined by using data from all other cycles except the $i$:th cycle.
Therefore, the fitted relationships between the parameters and precursors (Eqs. \ref{eq_BA_odd}, \ref{eq_CA_odd}, \ref{eq_even1}, \ref{eq_even2}, \ref{eq_even3}, \ref{eq_GMA_prec}) change from one predicted cycle to the next. 
This variability in the model parameters together with the residual variability incorporates the total prediction uncertainty of the model. It is important to note that the numerical values in Eqs. \ref{eq_even1}, \ref{eq_even2} and \ref{eq_even3} for
$A$, $B$ and $C$ of even cycles and in Eqs. \ref{eq_GMA_prec}, \ref{eq_BA_odd} and \ref{eq_CA_odd} for $A$, $B$ and $C$ of odd cycles as well as their standard errors
correspond to the fitted values when all available sunspot cycles are used in the fit. These values are therefore appropriate for prediction of sunspot cycles from cycle 25 onward. 

An important step in applying the cross-validation method is to estimate the prediction error of the model. Therefore, when proceeding through all the past cycles 1-24 (excluding odd cycles 1, 3, 5, 7 and 9, for which no geomagnetic precursor could be determined) we obtained the residuals of the 13-month smoothed SSN$^{3/4}$ values and the corresponding predicted values each time neglecting the cycle to be predicted as
\begin{equation}
\label{eq_res}    r = \textnormal{SSN}^{3/4} - \textnormal{SSN}_{\textnormal{pred}}^{3/4}.
\end{equation}
The exponent of $3/4$ in the above equation was used because the residuals calculated this way were more homoscedastic, i.e., the residual variance
was more uniform over different values of SSN compared to regular residuals in linear scale (exponent of 1 in Eq. \ref{eq_res}).
We then used the collection of residuals from Eq. \ref{eq_res} to generate a spread of possible sunspot number predictions for the $i$:th predicted cycle. This was done by bootstrapping $10^5$ values for the residuals (i.e., randomly resampling with replacement from the collection of residuals)
for each predicted monthly sunspot number value in the cycle. The residuals were added to the SSN$_{\textnormal{pred}}^{3/4}$ values and the result was then converted back to linear scale by the transformation
\begin{equation}
    \label{eq_res_inv}    \textnormal{SSN}_k = \left(\textnormal{SSN}_{\textnormal{pred}}^{3/4} + r_k\right)^{4/3},
\end{equation}
where $k=1,2,...,10^5$, $r_k$ indicates the $k$:th bootstrapped residual and $\textnormal{SSN}_k$ indicates the $k$:th predicted SSN value for a particular month. These $10^5$ values form an ensemble of SSN predictions for each monthly SSN value in the predicted cycle.

\begin{figure*}[htbp]
\includegraphics[width=\textwidth]{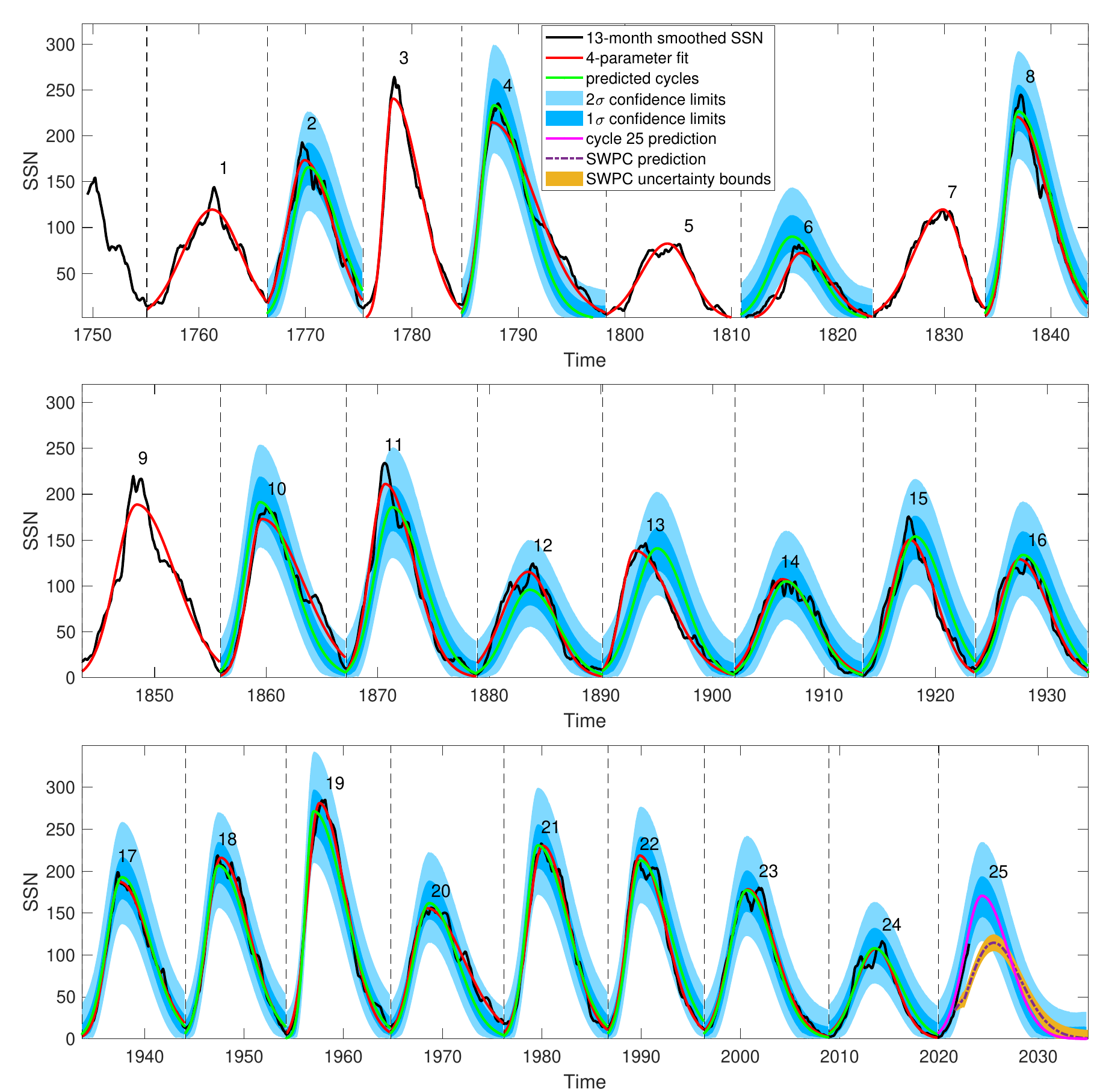}
\caption{Predicted past sunspot cycles as well as the prediction for cycle 25. The black curve shows the original 13-month smoothed SSN.
The red curves show the optimal 4-parameter fits to the SSN cycles. The green curves show the sunspot cycles predicted using the precursor values.
The dark and light blue shadings show the 1- and 2-standard deviation ranges based on the ensemble of SSN predictions. The magenta curve shows the prediction for cycle 25 and the dashed purple curve (and yellow uncertainty range) shows the cycle 25 prediction according to the international sunspot prediction panel obtained from the Space Weather Prediction Center.}
\label{fig9}
\end{figure*}

Figure \ref{fig9} shows the predicted past cycles (green curves) and their uncertainty ranges (blue shading) together with the 13-month smoothed
sunspot number (black curve) and the optimal 4-parameter model curves (red curves) as in Figure \ref{fig1}. The figure also shows 
the predicted curve for cycle 25 (magenta curve) and as a comparison the cycle 25 prediction offered by the Space Weather Prediction Center (https://www.swpc.noaa.gov/products/solar-cycle-progression). One can see that the predicted past cycles are rather close to both the optimal 4-parameter model curves and the 13-month smoothed sunspot number. Particularly noteworthy is the fact that the amplitude of all of the 
predicted past cycles quite accurately matches the real amplitude of the sunspot cycles. This is true for the very small cycles 6, 12, 14 and 24 and also for the largest cycles, e.g., cycle 19. Furthermore, by definition about 95\% of the monthly values of the 13-month smoothed sunspot number are within the 2-standard deviation uncertainty range of the predicted curve. This extremely good performance of the past predictions gives 
strong reasons to believe that the prediction of the sunspot cycle 25 is reliable as well.

\begin{figure*}[htbp]
\includegraphics[width=\textwidth]{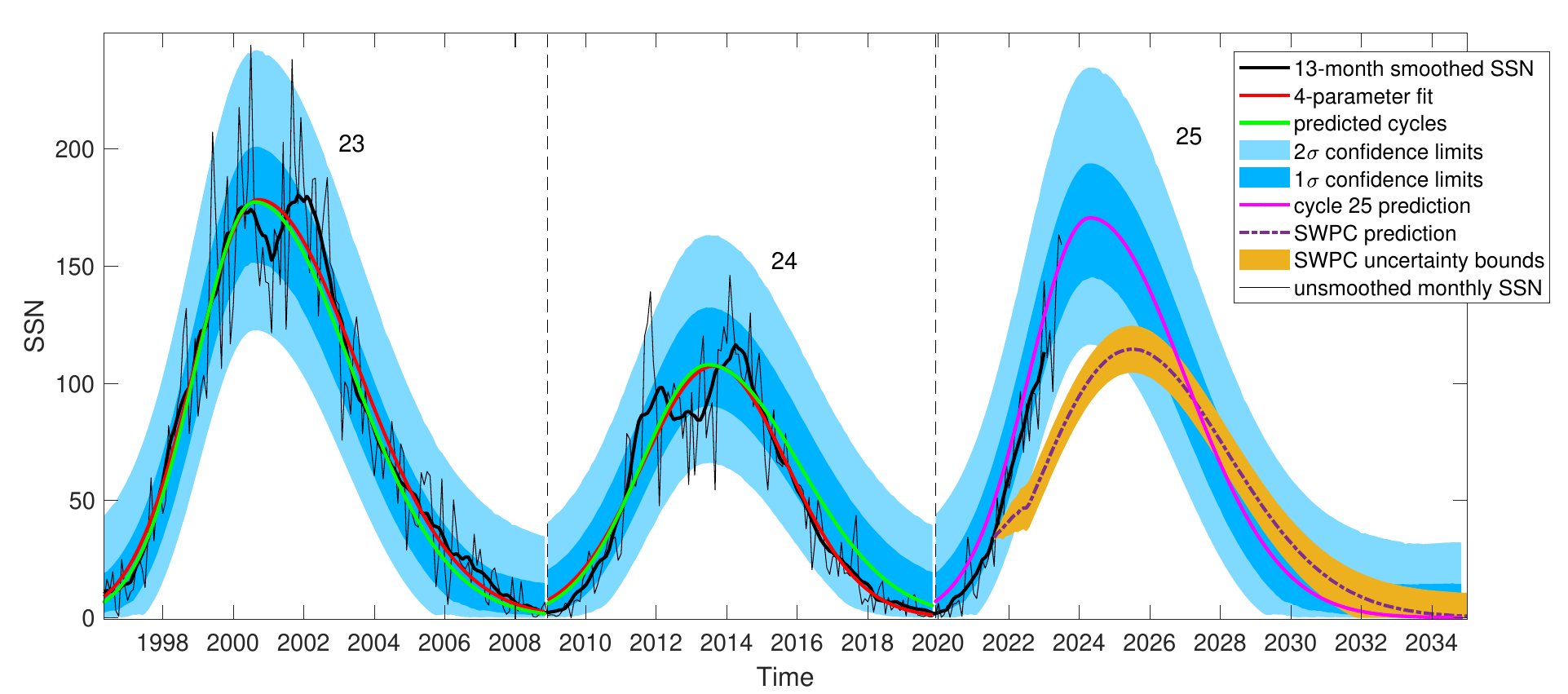}
\caption{Closeup of the sunspot prediction for cycle 25. The thick (thin) black curve shows the original 13-month smoothed (unsmoothed monthly) SSN. The red curves show the optimal 4-parameter fits to the SSN cycles. The green curves show the sunspot cycles predicted using the precursor values.
The dark and light blue shadings show the 1- and 2-standard deviation ranges based on the ensemble of SSN predictions. The magenta curve shows the prediction for cycle 25 and the dashed purple curve (and yellow uncertainty range) shows the cycle 25 prediction according to the international sunspot prediction panel obtained from the Space Weather Prediction Center.}
\label{fig10}
\end{figure*}

Figure \ref{fig10} shows a closeup of cycle 25 prediction in the same format as in Figure \ref{fig9}, but with the addition of the monthly 
unsmoothed values of SSN also included as the thin black curve. Our predicted curve peaks in May 2024 and the peak value of the cycle is $171\pm23$ (1 standard deviation uncertainty). 
Using Eq. \ref{eq_cycle_length} and the $D$ and $C$ parameters of cycle 24 the predicted length of cycle 25 is about $9.7\pm1.9$ years (2 standard deviation confidence interval).
This indicates that the end of cycle 25 is attained likely in September 2029 with the 2-standard deviation uncertainty range extending from October 2027 to July 2031.

At the time of writing this paper the cycle 25 has already begun and the recorded SSN is already significantly higher than 
the prediction offered by the SWPC. On the other hand, the recorded SSN follows quite closely our predicted curve, which indicates that the 
cycle 25 will be considerably stronger than cycle 24 and roughly the same size as cycle 23.
There have also been a range of other predictions for cycle 25. For example, \cite{Upton2018} used advective flux transport modeling
to predict that cycle 25 amplitude will be about 95\% of cycle 24 amplitude which would make cycle 25 the weakest cycle in 100 years.
\cite{Bhowmik2018}, on the other hand, used solar dynamo modeling to predict that cycle 25 would be slightly stronger (about 14\%) than cycle 24.
\cite{Pesnell2018} used the SODA index (based on a combination of solar polar magnetic field and solar F10.7 index) to predict
an amplitude of 135$\pm$25 for cycle 25, i.e., slightly larger than cycle 24. \cite{Kumar2021} used various polar field precursors extracted,
e.g., from solar magnetograms, to predict a cycle amplitude of 126$\pm$3 for cycle 25. \cite{Kumar2022} used the correlation between the rise rate
of polar field and the amplitude of the next cycle to predict an amplitude of 137$\pm$23 for cycle 25.
\cite{Sarp2018} used non-linear time series modeling approach to predict a clearly stronger cycle 25 with an amplitude of 154$\pm$12 peaking in early 2023. 
\cite{Li2018} used statistical modeling to reach a similar prediction with a predicted amplitude of 168.5$\pm$16.3 and peak of the cycle in October 2024. Both \cite{Sarp2018} and \cite{Li2018} predictions are fairly close, but slightly lower than our prediction.
A quite different prediction was given by \cite{McIntosh2020}, who used the timing of termination of toroidal bands of solar activity
to predict a rather strong cycle 25 with amplitude of 233$\pm$21. However, this prediction was recently revised to an amplitude of 184$\pm$17 \citep{McIntosh2022}, which is in agreement with our prediction when considering the ranges of uncertainty. \cite{Du2022b} used the correlation between the cycle rising rate and cycle amplitude to predict the cycle 25 based on the 2 years of data from the beginning of the cycle. They predicted the cycle amplitude to be 135.5$\pm$33.2, which is somewhat lower than our prediction, although not in complete disagreement given the uncertainty range. \cite{Penza2021} found a correlation between the parameters describing the sunspot cycle shape of even and subsequent odd cycles. Based on this they predicted the cycle 25 to be similar or slightly larger than cycle 24.
While the studies discussed above are only a subset of all predictions made for cycle 25 it seems that our prediction 
is clearly above the average in predicted cycle amplitude and also clearly above the SWPC prediction issued by the Solar Cycle Prediction Panel.

\subsection{Attempt at predicting the cycle 26}

Above we found that for even numbered cycles the 13-month smoothed sunspot number evaluated 41 months before the start of cycle provides an extremely good
estimate for the amplitude and other parameters of the cycle. This leads to an interesting question: how accurately could one use the predicted cycle 25 SSN curve 
to provide a prediction of cycle 26? Evidently the uncertainty of such a prediction would be rather large but we shall here attempt to make one also for 
cycle 26. The first step is to evaluate how well the predicted SSN 41 months before the sunspot minima actually correspond to the true values, which were used
as a precursor for even cycles. 
\begin{figure*}[htbp]
\includegraphics[width=\textwidth]{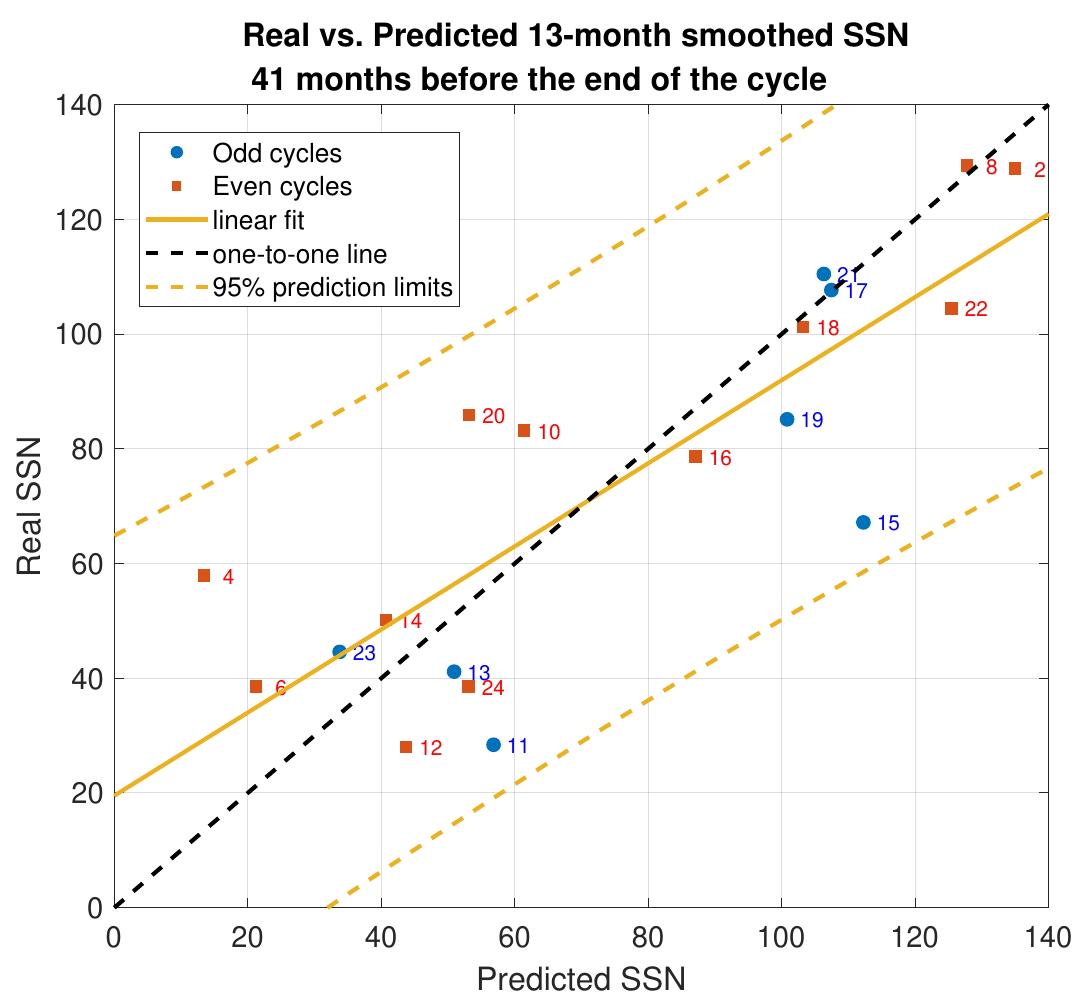}
\caption{Relationship between the real and predicted 13-month smoothed SSN evaluated 41 months before the minima that end each sunspot cycle.
The blue dots (red squares) indicate odd (even) cycles and the yellow line is a linear fit to all cycles.}
\label{fig11}
\end{figure*}

Figure \ref{fig11} shows the relationship between the real and predicted 13-month smoothed SSN evaluated 41 months before the minima 
that end each sunspot cycle. Both odd and even cycles seem to adhere to the same overall linear relationship, which has a high correlation of 0.832 (p $=10^{-5}$).
The linear relationship is given by the equation
\begin{equation}
       \textnormal{SSN(41)}_{\textnormal{real}} = 20(\pm10) + 0.72(\pm0.12)\times\textnormal{SSN(41)}_{\textnormal{pred}}.
       \label{eq_SSN41}
\end{equation}
Ideally there would be one-to-one relationship between the real and predicted values, but the modeled SSN curves seem to systematically slightly underestimate (overestimate) 
small (large) SSN(41) values. We can use this fit and its 95\% prediction error limits (see Figure \ref{fig11}) to predict the SSN 41 months before the sunspot minimum that
starts cycle 26. The timing of this minimum is determined by the timing of the minimum that starts cycle 25 and the cycle 25 length evaluated above from Eq. \ref{eq_cycle_length}.
Once the SSN(41) value to be used as a precursor for cycle 26 is known we can use Eq. \ref{eq_even1} to estimate the cycle amplitude.

\begin{figure*}[htbp]
\includegraphics[width=\textwidth]{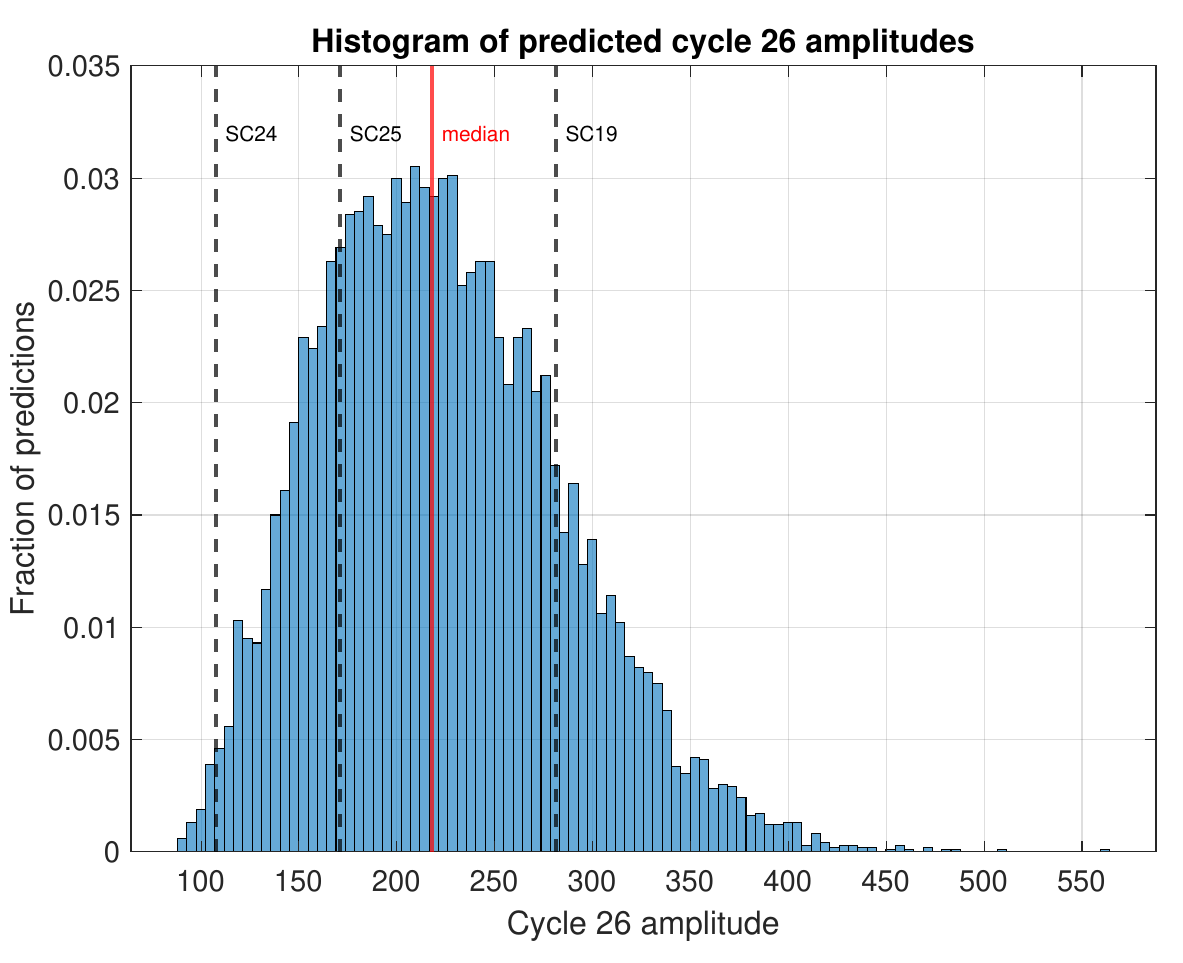}
\caption{Histogram of predicted cycle 26 amplitudes based on a Monte Carlo simulation with $10^4$ repetitions. For reference the dashed vertical lines show the cycle amplitudes
for cycle 19, 24 and predicted 25. The red dashed line indicates the median of the simulated values.}
\label{fig12}
\end{figure*}

Because of the uncertainties associated to the length of cycle 25 and the scaling of predicted SSN(41) according to Eq. \ref{eq_SSN41} we evaluated the spread of 
possible cycle 26 amplitudes using a Monte Carlo simulation having $10^4$ rounds. In each round we randomly generated a value for the length of the cycle 25 within the range 
of its uncertainty. This was then used to calculate the timing of 41 months before the end of cycle 25 and the SSN value at that time using the predicted SSN curve for 
cycle 25 (Figure \ref{fig10}). This value was then used to calculate a prediction for the cycle amplitude using Eq. \ref{eq_even1}.
The histogram of the Monte Carlo simulation results is shown in Figure \ref{fig12}. As expected the results indicate a quite a large uncertainty range covering practically
all past sunspot cycles. However, despite the large range of uncertainty some interesting and non-trivial aspects are seen. The median of the results implies that cycle 
26 would be even slightly stronger than cycle 25. In fact, based on these results the probability that cycle 26 will be weaker than cycle 25 is only about 19\%.
The results also imply an even clearer difference to cycle 24, which was the weakest cycle of the last 100 years. The probability that cycle 26 would be weaker than cycle 24 is only
0.8\%.

\section{Discussion and conclusions}

We have studied here the prediction of sunspot cycles paying particular emphasis on the differences between odd and even numbered cycles.
Our approach was to first parameterize the sunspot cycle as an asymmetric 4-parameter Gaussian function, which was found to represent each sunspot cycle extremely well.
Several of the parameters were found to correlate, which allowed us to simplify the representation of the sunspot cycles.
Interestingly it was found that the correlations between the cycle parameters were significantly stronger in even cycles compared to the odd ones.
For example the well-known Waldmeier effect that cycle amplitude and its rise time are inversely correlated was more strictly valid for even cycles. A similar finding for the Waldmeier effect was also reported, e.g., by \cite{Du2022} and \cite{Dikpati2008}.
This result implies that for some reason the even numbered cycles possess a smaller dimensionality than odd numbered cycles.
Such a systematic difference between even and odd cycles may imply a connection to the more fundamental 22-year magnetic Hale cycle of the Sun. 
\cite{Du2022} also found significant differences in the Waldmeier effect between the two hemispheres with the southern hemisphere
displaying the normal Waldmeier effect, which is stronger in even cycles, while the northern hemisphere displayed an inverse-Waldmeier effect,
which is stronger in odd cycles. This implies that there may also be a connection between the Waldmeier effect and the hemispheric asymmetry 
in the sunspot number.

Our approach to prediction of sunspot cycles is based on statistical precursors for the cycle parameters. Following \cite{Cameron2007} we found that the 
13-month smoothed sunspot number evaluated 41 months before the sunspot minimum that begins a cycle, SSN(41), is on average a fair predictor for the next cycle when considering
all sunspot cycles. However, we found that this relation is much stronger for even cycles than for odd cycles, for which the SSN(41) could not be very accurately 
used as a predictor.
The advantage of SSN(41) precursor is that we can predict for even cycles all the cycle parameters (which correlate well with the amplitude) quite accurately.
\cite{Cameron2007} explained the tendency of past SSN to correlate with the next cycle amplitude as a result of overlapping sunspot cycles. 
While the past cycle is still declining the new cycle begins. Because of the Waldmeier effect the stronger the new cycle will be the faster it will rise to the maximum
and the earlier the intersection time of the old and the new cycle is attained. The earlier the intersection time, the higher the SSN 41 months earlier in the
declining phase of the previous cycle. The fact that this connection is here found to be more strictly valid for even cycles arises because of the fact that 
the Waldmeier effect is also tighter in even cycles. The question of why the Waldmeier effect is different in even and odd cycles is still open and should be investigated in future studies.

For odd cycles we found that the maximum Sep-Oct geomagnetic $aa$ index within 3 years 
preceding the sunspot minimum that begins the cycle is an extremely good precursor for the cycle amplitude. 
It has been long recognized that geomagnetic activity close to solar minimum reflects the strength of the open solar flux, which is at these times connected
to the poloidal magnetic flux extending from the Sun's polar coronal holes. It is curious, however, that the best precursor for odd cycles was found
to be connected to geomagnetic activity close to the fall equinox. Some other past studies have used geomagnetic activity averaged in other some ways over different periods
of time and found good correlations to the next cycle amplitude. While our Sep-Oct $aa$ precursor is probably not statistically significantly better than some of 
these other measures there might be a physical reason for the preference of Sep-Oct season. 

It is known that geomagnetic activity has strong seasonal variation, which is largely due to the Russell-McPherron (RMP) effect \citep{Russell1973,Lockwood2020}. 
The RMP effect describes the fact that interplanetary magnetic field (IMF), which is oriented along the solar equatorial plane, projects onto the Z-axis of the GSM
coordinate system close to the fall and spring equinoxes and thereby may enhance energy input from the solar wind into the magnetosphere/ionosphere system.
The RMP effect is also dependent on the polarity of the IMF so that during fall (spring) equinox the IMF pointing away from (towards) the Sun enhances 
solar wind energy input and therefore leads into larger geomagnetic activity as well. Often both polarities of the IMF are seen within a solar rotation but
especially close to solar minima the heliospheric current sheet is often flatter, which allows Earth to be more exposed to the magnetic polarity connected 
to Sun's northern (southern) pole in fall (spring) due to the Earth's changing heliographic latitude over the course of the year \citep{Rosenberg1969}.
According to the 22-year Hale cycle the northern solar pole has a positive magnetic polarity close to sunspot minima preceding odd cycles, 
which leads to a dominance of away sector close to fall equinoxes \citep{Hiltula2007,Vokhmyanin2012}. Furthermore, it has been shown 
that the heliospheric current sheet is systematically tilted southward in the declining phase \citep{Hiltula2006}, which further enhances the 
dominance of the away sector in fall and decreases the dominance of the toward sector in spring prior to odd cycles.
Therefore, it is expected that also the geomagnetic activity portrayed by the $aa$ index is most sensitively proportional to the strength of the IMF (i.e., to open solar magnetic flux connected to solar polar field)  at these times.
A detailed confirmation of this interpretation is warranted, but out of the scope of this study.

In addition to cycle amplitude and other parameters we found a curious statistical relation between the length of the sunspot cycle and the ratio of $D$ (cycle asymmetry) and $C$ (time scale of rising phase) of the preceding cycle. The fact that such properties of the preceding cycle somehow are connected to the length of the next cycle again 
highlights the fundamental 22-year Hale cyclicity of solar magnetism. While there is no physical explanation for this relationship at the moment 
it can be statistically used to estimate the cycle length perhaps a bit more accurately than previous metrics \citep{Petrovay2020}.

Using the found precursors we used cross-validation to test their prediction accuracy by predicting the past solar cycles. For all past cycles the predictions were 
very close to the real sunspot cycles thereby giving strong confidence that the prediction of future cycles would be equally successful.
We proceeded to predict the odd cycle 25 and found that its amplitude will be $171\pm23$, thus about 1.6 times stronger than cycle 24. There are already clear indications that our prediction
closely follows the progression of the cycle 25 which has already started at the time of writing this paper. It is also noteworthy that the prediction issued by the
Solar Cycle 25 Prediction Panel at the SWPC suggests cycle 25 to be similar to cycle 24, which is already now clearly below the current sunspot levels.

Using the predicted cycle 25 and the fact that SSN(41) could be used as a predictor for the even cycles we provided a rough prediction also for cycle 26.
As expected the uncertainty range of the prediction was rather large, but based on the results it seems rather likely that the cycle 26 will be stronger than 
both cycles 24 and 25. Therefore, we find no evidence for an imminent drop of solar activity to a grand solar minimum as suggested by several past studies \citep[e.g.][]{Abreu2008,Owens2011,Lockwood2011}.
Overall these results display the capability to predict even and odd cycles using different precursors with rather high accuracy. 
The results also clearly indicate a connection between odd-even pairs of sunspot cycles and highlight the 22-year Hale cyclicity. Accordingly,
the Hale cyclicity should be considered more carefully also by more physically motivated dynamo and flux transport prediction models of solar activity.

\begin{acknowledgements}
We acknowledge the financial support by the Academy of Finland to the PROSPECT (project no. 321440).
The sunspot number data was obtained from World Data Center SILSO, Royal Observatory of Belgium, Brussels (https://www.sidc.be/SILSO/home).
\end{acknowledgements}

%%% BIBLIOGRAPHY %%%%%%%%%%%%%%%%%%%%%%%%%%%%%%%%%%%%%%%%%%%%%%%%%%%%%%%%%%%

     % format of references provided by the journal (.bst)
%\bibliographystyle{aa}
     % name your Bibtex file containing your references (.bib)
\bibliography{references}

\end{document}